\documentclass[twocolumn,showpacs,preprintnumbers,amsmath,amssymb,superscriptaddress]{revtex4}
\usepackage{graphicx}% Include figure files
\usepackage{dcolumn}% Align table columns on decimal point
\usepackage{amsmath}    % need for subequations
\usepackage{verbatim}   % useful for program listings
\usepackage{color}      % use if color is used in text
\usepackage{subfigure}  % use for side-by-side figures
\usepackage{hyperref}   % use for hypertext links, including those to external documents and URLs
\usepackage{amsfonts}
\usepackage{bm}% bold math

\begin{document}

%\preprint{Toshiba Research Europe Limited/Confidential}

\title{ Polarization correlated photons from a positively charged quantum dot
}% Force line breaks with \\

\author{Y. Cao}
\affiliation{Toshiba Research Europe Limited, Cambridge Research Laboratory,\\
208 Science Park, Milton Road, Cambridge, CB4 0GZ, U. K.}
\affiliation{Department of Physics, Imperial College London, Prince
Consort Road, London SW7 2AZ, U. K.}

\author{A. J. Bennett}
\email{anthony.bennett@crl.toshiba.co.uk}
\affiliation{Toshiba Research Europe Limited, Cambridge Research Laboratory,\\
208 Science Park, Milton Road, Cambridge, CB4 0GZ, U. K.}

\author{I. Farrer}
\affiliation{Cavendish Laboratory, Cambridge University,\\
J. J. Thomson Avenue, Cambridge, CB3 0HE, U. K.}

\author{D. A. Ritchie}
\affiliation{Cavendish Laboratory, Cambridge University,\\
J. J. Thomson Avenue, Cambridge, CB3 0HE, U. K.}

\author{A. J. Shields}
\affiliation{Toshiba Research Europe Limited, Cambridge Research Laboratory,\\
208 Science Park, Milton Road, Cambridge, CB4 0GZ, U. K.}

\date{\today}%

\begin{abstract}
Polarized cross-correlation spectroscopy on a quantum dot charged
with a single hole shows the sequential emission of photons with
common circular polarization. This effect is visible without
magnetic field, but becomes more pronounced as the field along the
quantization axis is increased. We interpret the data in terms of
electron dephasing in the $X^{+}$ state caused by the Overhauser
field of nuclei in the dot. We predict the correlation timescale can
be increased by accelerating the emission rate with cavity-QED.
\end{abstract}

%\pacs{78.67.-n, 85.35.Ds}% PACS, the Physics and Astronomy

\maketitle %this tag has to go at end of title section

%Letter text 1,500 words, excluding the introductory
%paragraph, Methods, references and figure legends., 1st para 150 words

%For nature journals use:
%\textbf{ 1st paragraph text in bold }

%TO COMPILE. click Latex once, then pdf-latex then acrobat symbol.

Spins in quantum dots (QDs) provide a promising platform for
manipulating and storing quantum information in the solid state.
Optical measurements have demonstrated spin preparation
\cite{Atature06, Gerardot07} coherent spin control \cite{DeGreve11}
and electron-spin $-$ photon entanglement \cite{DeGreve12, Gao12}.
There are also proposals for achieving photon entanglement
\cite{Hu07} and non-destructive measurement of photons
\cite{Witthaut12} using charged QDs. However, the time evolution of
the carrier spin is unavoidably affected by the 10$^{4}$ $-$
10$^{5}$ nuclei in the dot, all with non-zero spin. One example of
the utility of the electron-nuclear interaction is its use in spin
pumping the hole into the spin down state in zero external field, by
pumping on the spin up state of $X^{+}$ \cite{Gerardot07}. From a
fundamental point of view then, the hyperfine interaction provides
an interesting system for manipulating a mesoscopic nuclear ensemble
and observing its dynamics.

It has now been established that the electron-nuclear hyperfine
interaction is dominated by the contact interaction and is isotropic
in a QD \cite{Urbaszek13}. The dynamics of this interaction
manifests itself in studies of polarized photo-luminescence
\cite{Cortez02, Laurent06}. Contrastingly, the hole's $p$-like
wave-function has a node at each nucleus leaving the dipole-dipole
interaction between the hole and nuclear spins to dominate
\cite{Testelin08}. This interaction has a strength one order of
magnitude below that of the electron \cite{Chekovich11, Fallahi10}.
Thus, there has been interest in using the hole-spin as quantum bit
with reduced decoherence. Direct measurements of the hole spin
relaxation time in a vertical magnetic field, $T_{1}^{h}$, have
shown it is hundreds of microseconds \cite{Heiss07, Gerardot07}.
Without applied magnetic field some experiments suggest the hole
spin $T_{1}^{h}$ time is 13 ns \cite{Urbaszek13}. Several studies
have now estimated the hole dephasing time, $T_{2}^{h}$ in magnetic
field is approximately 1 $\mu s$ \cite{Brunner09, DeGreve11}.

\begin{figure}[h]
\includegraphics[width=80mm]{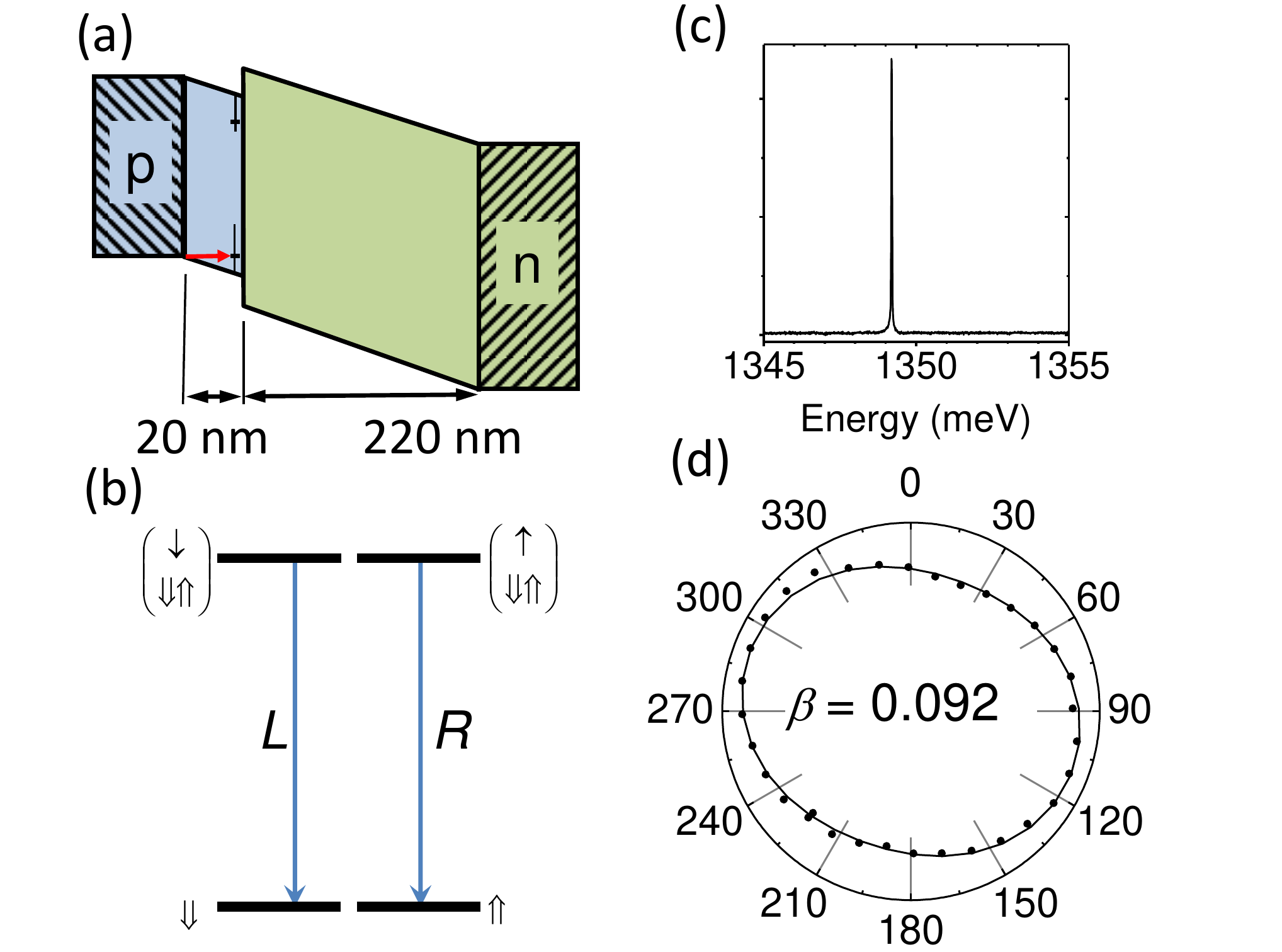}% Here is how to import EPS art
\caption{\label{Fig1} (a) Schematic band-structure of the
hole-charging diode when the dot contains a single hole. (b) Energy
level diagram for the $X^{+}$ at zero magnetic field. (c) Spectrum
under quasi-resonant excitation at 1.2 V. (d) Emission pattern for
an $X^{+}$ transition showing $\beta =$ 0.092. }
\end{figure}

We study here the emission from the $X^{+}$ state in a dot
deterministically charged with a hole using a diode (Figure
\ref{Fig1}a). We show photons from this transition display
polarization correlation over a timescale one order of magnitude
greater than the radiative lifetime when excited by a linearly
polarized laser. This time is short relative to some reports of the
hole spin polarization lifetime \cite{Heiss07} and we show that this
is a result of dephasing caused by the electron when the system is
excited. We investigate the magnitude of the effect as a function of
applied external magnetic field and radiative lifetime.

The $X^{+}$ consists of two holes in a singlet $S=0$ state and an
electron. The zero net-hole-spin ensures the electron-hole
anisotropic exchange interaction is absent. In zero magnetic field,
the $X^{+}$ eigenstates are degenerate and labeled by the spin of
the electron (Figure \ref{Fig1}b) \cite{Bayer1999B, Bayer99}.
Radiative decay from $X^{+}$ to the single hole ground state occurs
with a change in total angular momentum $\pm 1 $, the polarization
of the photon being correlated with the initial and final spin
state. In a dot where the hole is ``heavy" ($m_{j} = \pm 3/2$) only
the vertical transitions in Figure \ref{Fig1}b are allowed: the
detection of a left-handed photon (L) photon ensures the decay
occurred by the left hand transition on Figure \ref{Fig1}b.

The strain, shape anisotropy and inversion asymmetry in InGaAs/GaAs
QDs ensures the optically active transition has a mixed heavy-
($m_{j} = \pm 3/2$) and light- ($m_{j} = \pm 1/2$) hole character
\cite{Koudinov04, Belhadj10}. The state is given by $\phi_{\pm} =
(|\pm 3/2 \rangle + \beta |\mp 1/2 \rangle )/ \sqrt(1+\beta^{2})$,
which we denote $\Uparrow$ and $\Downarrow$. Recombination of an
electron and a mixed heavy/light-hole now results in elliptically
polarized photons from $X^{+}$, and $\beta$ may be determined from
the emission pattern (Figure \ref{Fig1}d) \cite{Koudinov04}). Within
the sample studied $\beta$ values from 0.02-0.20 are typical, and
for the data shown here $\beta= $0.092 (Figure \ref{Fig1}d).

\begin{figure}[h]
\includegraphics[width=80mm] {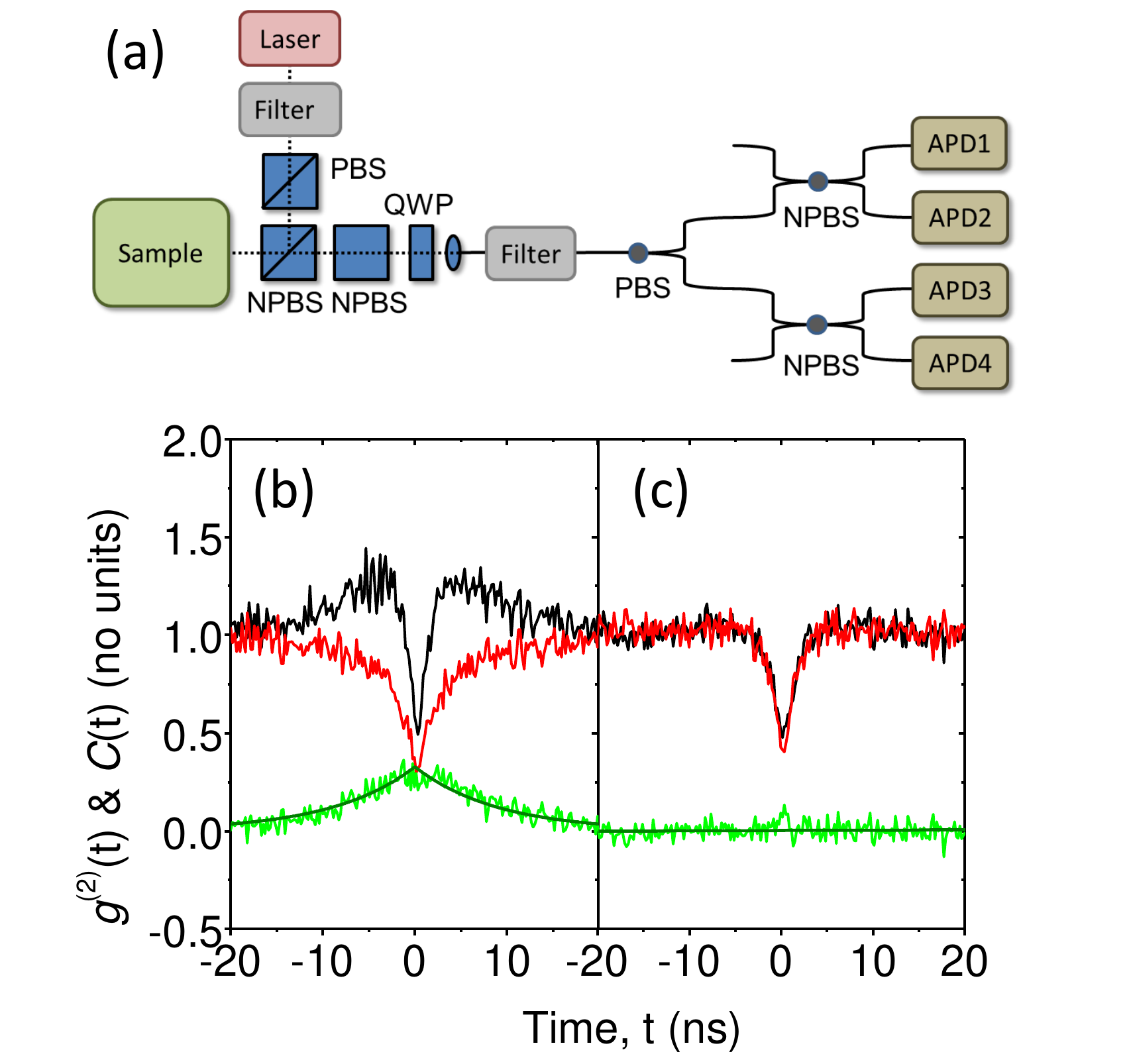}
\caption{\label{Fig2} (a) Apparatus to measure the polarization
correlation from a dot. (b) Circular co- and cross- polarized
emission correlation from $X^{+}$ at zero external field,
$g_{co}(t)$ (black) and $g_{cross}(t)$ (red), respectively.
Extracted degree of polarization correlation, $C(t)$ (green). (c)
The same measurement made in the linear detection basis.}
\end{figure}

The diode for controlled charging has a 20 nm GaAs tunnel barrier
between the dot and $p$-contact (Figure \ref{Fig1}a). A 75\% AlGaAs
barrier on the $n$-side prevents electron charging, so the $X^{+}$
dominates at 1.2-1.3 V. Emission from $X^{+}$ at an energy of 1349.2
meV is excited quasi-resonantly by a linearly polarized laser at
1317.2 meV. This excitation scheme equally excites both transitions
in Figure \ref{Fig1}b and the absence of spin pumping ensures there
is no build up of nuclear spin polarization, but it does not
populate other carrier combinations such as neutral and negatively
charged excitons (Figure \ref{Fig1}c).

The experiment is shown in Figure \ref{Fig2}a. After filtering, the
emission is passed to polarization-maintaining fibre-optics which
enable 4 simultaneous measurements of correlation in a basis
selected by the quarter-wave plate (QWP) and the polarizing coupler
(PBS).

Figure \ref{Fig2}b presents correlations recorded at zero external
field at an excitation power $\times$10 below saturation, in the
circular basis. Comparing the sum of the two co- and cross-polarized
measurements ($g_{co}(t)$ in black and $g_{cross}(t)$ in red,
respectively), we see a clear difference. Note that both $g_{co}(t)$
and $g_{cross}(t)$ show a reduced signal within $\sim$ 1 ns of
zero-time delay due to the anti-bunched nature of the light. Outside
the central $\sim$ 1 ns there is an enhanced probability of the
source emitting two photons of the same circular polarization over
the case of emitting photon of opposite circular polarization.

The degree of polarization correlation, $C(t)$ is defined as $C(t) =
(g_{co}(t)-g_{cross}(t))/(g_{co}(t)+g_{cross}(t))$ from which a
least squares fit with a function $C(t) = C_{0}*exp(-|t|/ \tau_{d})$
 extracts the polarization correlation at zero delay
$C_{0}$ and the timescale, $\tau_{d}$. Empirically, this function is
a good fit to $C(t)$ (Figure \ref{Fig2}b). $C_{0}= 0.33 \pm 0.01$
and the decay time of the correlation $\tau_{d} = 9.0 \pm 0.4ns$. In
contrast, measurements in the linear-polarization basis (H/V) show
an absence of polarization correlation (Figure \ref{Fig2}c).

Non-zero heavy-light hole mixing is an obvious source of reduced
polarization correlation. Taking the heavy: light hole oscillator
strength of 3:1 \cite{Belhadj10, Koudinov04} we see that
recombination of a $\phi_{+}$ hole and an electron in the $X^{+}$
level leads to an elliptical photon with state $\propto \sqrt{3}|L>
+ \beta |R> $, and a $\phi_{-}$ hole. Conversely, decay involving a
$\phi_{-}$ hole and an electron leads to a $\propto \sqrt{3}|R> +
\beta |L> $ photon. The measurement in Fig. \ref{Fig2}b is in the
circular basis so detection of a left-handed photon implies the
decay came from $\phi_{+}$ with $3/(3+\beta^{2})$ probability. In
the absence of dephasing in the upper or lower states, this reduces
the probability of obtaining sequential left-left photo-detections
to $(9+\beta^{4})/(3+\beta^{2})^{2}$. For this QD $\beta =$ 0.092,
so the probability of co-polarized photon emission is reduced to
0.994. This is higher than we have measured, so we conclude an
additional factor must be included.

\begin{figure}[h]
\includegraphics[width=60mm]{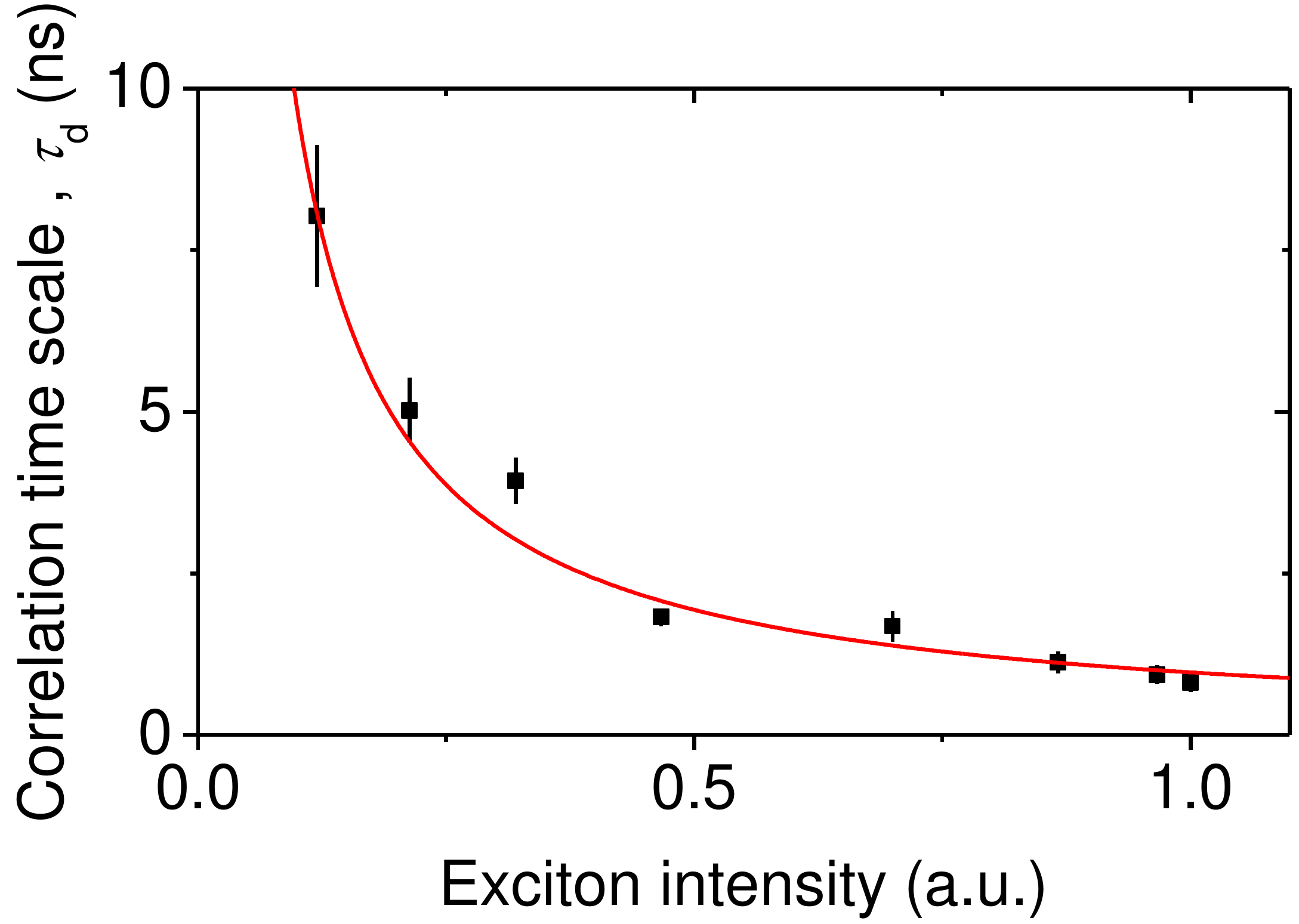}
\caption{\label{Fig2B} A measurement of the timescale of correlation
at zero external magnetic field, $\tau_{d}$, as a function of the
normalized intensity of the source. The data is fitted with an
inverse relationship.}
\end{figure}

In fact, the data can be explained by the fast dephasing of the
electron spin in the upper state, which dominates any dephasing from
the hole spin. A coincidence detection event arises as follows: the
transition emits a photon that is detected with circular
polarization and the hole spin is left in the corresponding state.
Some time later, the system is re-excited to the upper state, where
electron spin dephasing occurs during the radiative lifetime of the
$X^{+}$ state, following which a second photon is emitted from the
spontaneous decay. These two photons form a single coincidence in
Figure \ref{Fig2}b. We stress that our model implicitly assumes the
hole spin lifetime is greater than the measured $\tau_{d}$, though
we envisage that future experiments that reduce the effect of
electron dephasing it will be necessary to include the contribution
of the hole.

Our studies provide four pieces of evidence electron spin dephasing
is the factor limiting the polarization correlation. Firstly, the
degree of polarization correlation from the $X^{+}$ observed in
Figure \ref{Fig2}b is 1/3. When excited, the unpaired electron spin
evolves through a hyperfine interaction with the nuclei. Only those
nuclear field fluctuations in the two directions perpendicular to
the spin will cause precession. If this precession is faster than
the radiative lifetime of the upper state, its effect is to
randomize the spin. The electron spin parallel to the nuclear field
is preserved. Thus, the mean spin projection along \textbf{z} is
reduced to 1/3. Secondly, the timescale over which polarization
correlation is observed is inversely proportional to pump rate, as
shown in Figure \ref{Fig2B}. This cannot be explained by dephasing
occurring in the ground state. The increased excitation increases
the number of times the system is excited between photon detection
events, and this increases the rate at which the polarization
correlation is lost. Thirdly, there is no polarization correlation
in the linear basis (Figure \ref{Fig2}c). This is consistent with a
dephasing of the electron spin state in a time faster than the
$X^{+}$ radiative lifetime. Finally, we shall show that change in
$C_{0}$ with magnetic field is explained by the dynamics of the
electron spin.

\begin{figure}[h]
\includegraphics[width=80mm]{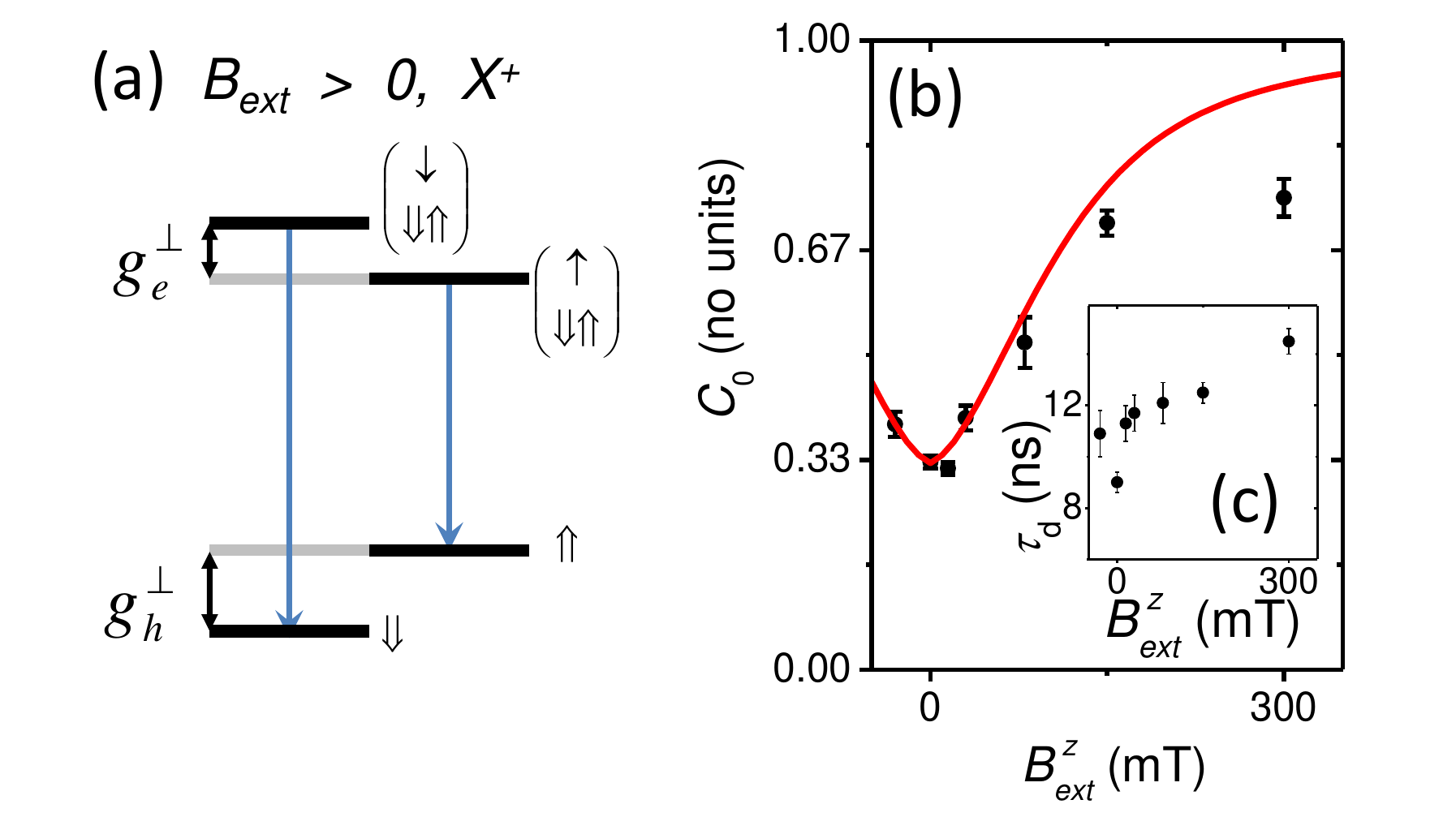}% Here is how to import EPS art
\caption{\label{Fig3} Energy level diagram for the $X^{+}$
transition with finite z-magnetic field. (b) The degree of
correlation, $C_{0}$ as a function of external Faraday magnetic
field (black points). The calculated variation with field is shown
as a red line. (c) Variation in the timescale of correlation,
$\tau_{d}$.}
\end{figure}

We next discuss the application of a Faraday magnetic field, which
removes the degeneracy of the upper and lower states, shown in
Figure \ref{Fig3}a. The net field experienced by the spins is the
sum of the external field $B_{ext}$ and the internal nuclear field,
$B_{N}$. This stabilizes the electron spin along \textbf{z} and
causes it to precess about the sum of the two fields, which is
predominantly along the \textbf{z} when the $|B_{ext}| > B_{N}$.
Thus, the application of vertical field increases the value of
$C_{0}$ as shown in Figure \ref{Fig3}b. Figure \ref{Fig3}c plots the
polarization correlation timescale, $\tau_{d}$, versus magnetic
field at constant laser intensity. This value changes from 9.0 $\pm$
0.4 ns at zero field to 14.5 $\pm$ 0.5 ns at 300 mT.

A model of the dephasing of electron spin in QDs was presented by
Merkolov, Efros and Rosen \cite{Merkolov02}. In this framework it is
assumed that on timescales below 1 $\mu s$ the hyperfine interaction
between the electron spin and the nuclei in the dot can be
considered semi-classically as a ``frozen" magnetic field, of finite
variance, but no directional preference. The electron $g$-factor is
assumed isotropic. The time evolution of the electron spin
$\mathbf{S}(t)$, (initially along $\mathbf{S}_{0}$) is given by:

\begin{subequations}
 \begin{equation}\label{equ1}
 \mathbf{S}(t) =
 \\
 (\mathbf{S}.\mathbf{n})\mathbf{n} + \{ \mathbf{S}_{0} -
 \\
 (\mathbf{S}_{0}.\mathbf{n}) \mathbf{n}\} cos(\omega t) +
  [\{\mathbf{S}_{0} -
 (\mathbf{S}_{0}.\mathbf{n})\mathbf{n} \}\times \mathbf{n}]sin(\omega t)
 \end{equation}
 \begin{equation}
 W(B_{N}) \propto exp[-\frac{(B_{N})^{2}}{\delta B_{N}^{2}}]
 \end{equation}

\end{subequations}

\begin{figure}[h]
\includegraphics[width=80mm] {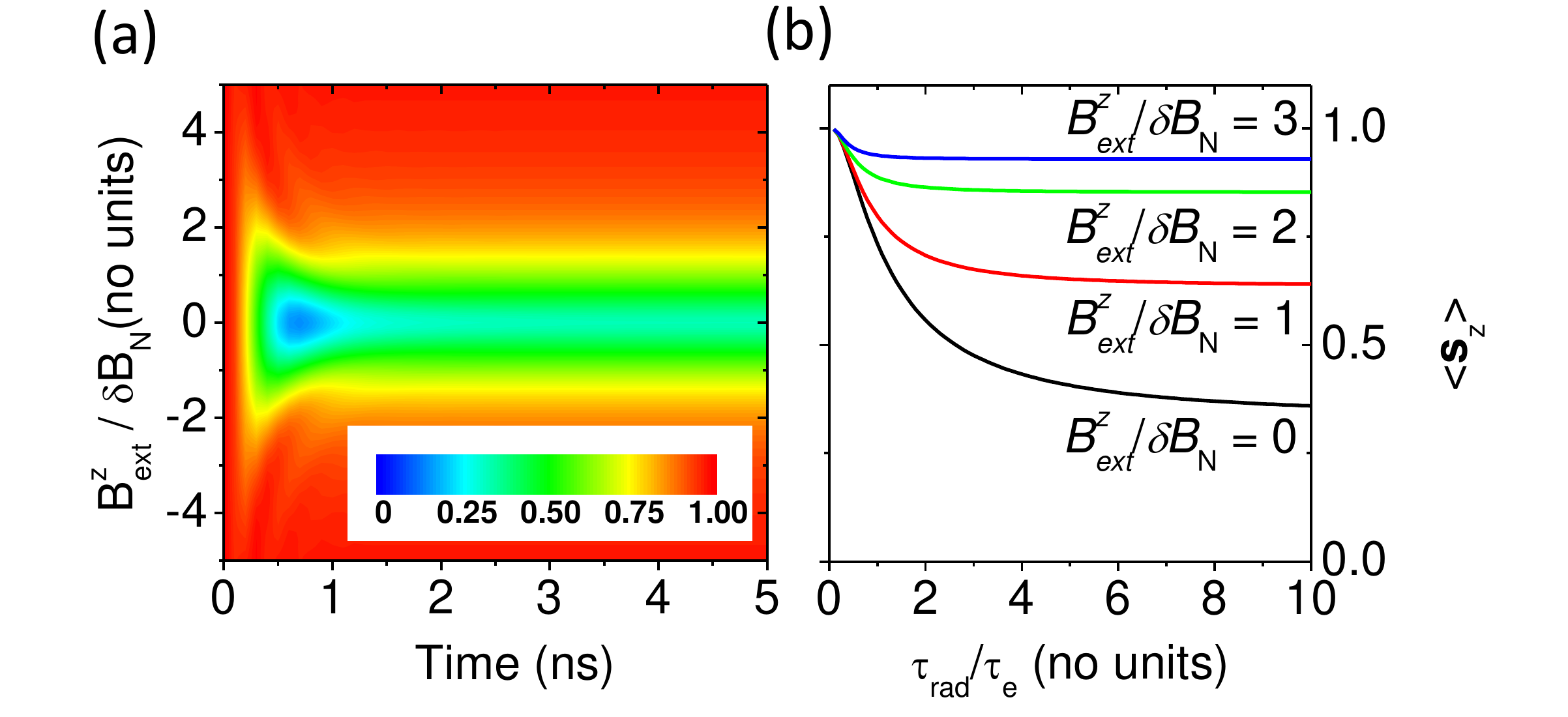}
\caption{\label{Fig4} (a) Projection of the electron spin along the
z-direction as a function of magnetic field and time, assuming a
nuclear field fluctuation of width $\delta B_{N} =$ 100 mT and an
electron $g$-factor of 0.5\cite{Bennett13}. (b) The averaged
projection of the z-component of the spin over the radiative
lifetime, scaled in terms of the electron spin dephasing time.}
\end{figure}

Where the distribution of nuclear field strengths, $W(B_{N})$ is
parameterized by the Gaussian width of fluctuations, $\delta B_{N}$
. Figure \ref{Fig4}a shows how the spin projection along the
\textbf{z} direction $S_{z}$ varies with the external magnetic field
$B^{z}_{ext}$ \cite{Merkolov02}. At fields of a few times $\delta
B_{N}$ the spin-projection along \textbf{z} is stabilized. This has
not eliminated the nuclear spin fluctuations, it has merely
overwhelmed them at the cost of increased rate of precession about
the net field. Any measurement along an orthogonal polarization
direction will reveal the increased precession rate.

From Equation \ref{equ1} we extract the expected final electron spin
projection along \textbf{z}, $S_{z}$, which is equal to the
polarization correlation $C_{0}$ (solid line in Figure \ref{Fig3}b).
The only fitting parameter is the width of the fluctuations in the
nuclear-field $\delta B_{N}$, set to 100 mT. The dephasing time for
this electron \cite{Urbaszek13} is therefore
$T_{\bigtriangleup}=\hbar/g_{e} \mu_{B} \delta B_{N} \sim 200 ps$,
which is, as expected, much less than the 1 ns radiative lifetime of
the upper state. This provides a good fit to the data, reproducing
the value of $C_{0}$ and width around zero field. The model fits
less well at the higher fields. Partly, this can be explained by
non-zero $\beta$, but the discrepancy requires further
investigation.

The extracted $\delta B_{N}$ is within the range derived from a
spin-noise measurement \cite{Kulhmann13} but is greater than
inferred from dephasing of the $X_{0}$ state in similar dots
\cite{Bennett11}. We attribute this to the smaller wave-function
extent of the electron when the dot additionally contains two holes.
As $\delta B_{N}$ scales with $1/ \sqrt(V)$ where $V$ is the volume
of spins overlapping with the wave-function there is a variation in
the effective $\delta B_{N}$ between states. This is the same reason
the electron $g$-factor changes in the presence of additional holes
\cite{Bennett13}.

To increase $C_{0}$ one could employ a host semiconductor without
nuclear spin, a QD of greater volume or reduce the fluctuations in
the nuclear field. Alternatively, Figure \ref{Fig4}b shows that
reducing $\tau_{rad}$ to the electron spin-lifetime leads to a
significant increase in polarization correlation. This could be
achieved by placing the dot into a cavity that equally enhances the
radiative decay, independent of polarization.

In conclusion, despite the long hole spin coherence time in quantum
dots the emission of polarization-correlated photons from the
$X^{+}$ state is limited by electron spin dephasing. A significant
increase in the polarization correlation time should be achieved by
reducing the radiative lifetime of the $X^{+}$ state. Additionally,
the degree of polarization correlation can be increased by applying
an external magnetic field greater than the nuclear field.

\section{Acknowledgements}

We thank the EPSRC  for funding the molecular beam epitaxy (MBE)
machine and the Controlled Quantum Dynamics Centre for Doctoral
Training (CQD-CDT) which supported Y. Cao.

% *****     References     *****
%*******************************

\end{document}